\documentclass[fleqn,preprint,showpacs,preprintnumbers,amsmath,amssymb]{revtex4}
\usepackage{graphicx}% Include figure files
\usepackage{dcolumn}% Align table columns on decimal point
\usepackage{bm}% bold math
\begin{document}

\title{Nonlinear Excitations in Strongly-Coupled Fermi-Dirac Plasmas}
\author{M. Akbari-Moghanjoughi}
\affiliation{Azarbaijan University of
Tarbiat Moallem, Faculty of Sciences, Department of Physics, 51745-406, Tabriz, Iran}

\date{\today}
\begin{abstract}

In this paper we use the conventional quantum hydrodynamics (QHD) model in combination with the Sagdeev pseudopotential method to explore the effects of Thomas-Fermi nonuniform electron distribution, Coulomb interactions, electron exchange and ion correlation on the large-amplitude nonlinear soliton dynamics in Fermi-Dirac plasmas. It is found that in the presence of strong interactions significant differences in nonlinear wave dynamics of Fermi-Dirac plasmas in the two distinct regimes of nonrelativistic and relativistic degeneracies exist. Furthermore, it is remarked that first-order corrections due to such interactions (which are proportional to the fine-structure constant) are significant on soliton dynamics in nonrelativistic plasma degeneracy regime rather than relativistic one. In the relativistic degeneracy regime, however, these effects become less important and the electron quantum-tunneling and Pauli-exclusion dominate the nonlinear wave dynamics. Hence, application of non-interacting Fermi-Dirac QHD model to study the nonlinear wave dynamics in quantum plasmas such as compact stars is most appropriate for the relativistic degeneracy regime.
\end{abstract}

\maketitle

\section{Theoretical Review}

Studies reveal that a degenerate plasma can behave much differently compared to the classical counterparts \cite{bohm, manfredi, shukla, haas1, Markowich, Marklund, Brodin}. This is mainly due to the quantum nature of particle interactions involved in this kind of plasmas. An idealized Fermi-plasma model is the one in which fermions have much lower energies compared to the characteristic Fermi energy defined by the total number of fermions. In such case the quantum plasma is referred to as the zero-temperature quantum plasma. Despite the name, i.e. zero-temperature, it has been shown that such model is useful in description of some compact stellar objects with relatively higher temperatures as $10^4K$ \cite{chandra1}. In a noninteracting low density degenerate Fermi gas model which is usually considered for solid-state materials the Pauli exclusion \cite{landau} is the dominant quantum effect although the quantum tunneling may also play a role in hydrodynamic properties of plasma. It has been shown that the tunneling effect which has negative pressure-like nature can lead to dissipation of nonlinear structures in quantum plasmas \cite{haas2, akbari1, akbari2}.

The main reason for increasing interest in the study of quantum plasmas is two-fold. Semiconductors, inertial confined dense plasmas, laser-mater interaction, etc. can be treated as quantum-like plasmas. On the other hand, astrophysical compact-stars such as dwarfs, pulsars, etc. can also be studied within the framework of quantum plasmas. The study of degenerate Fermi gas under extreme conditions such as ultra-high density and magnetic field can lead to a better understanding of internal structures and physical process in superdense astrophysical entities and stellar chain of evolution. Densities of the order $10^6$-$10^9gcm^{-3}$ may arise in white-dwarfs due to gigantic gravitational force leading to a final collapse \cite{chandra2, chandra3}. It is well known that the thermodynamical properties \cite{kothary} and nonlinear wave dynamics \cite{akbari3, akbari4, akbari5} of a degenerate plasma can exhibit distinct behavior in nonrelativistic and ultrarelativistic degeneracy limits. There may be a mechanism in some highly magnetic white-dwarfs to generate fields as high as $10^7G$. One of these mechanisms is called the Landau orbital ferromagnetism (LOFER) \cite{connel}. Recently, it has been shown that such mechanism can lead to a collapse in the internal transverse plasma pressure the state so-called quantum-collapse in white-dwarfs \cite{akbari6}. However, the study of quantum plasma under such extreme cases using the conventional noninteracting QHD models used for ordinary degenerate plasmas may not seem quite appropriate at the first place.

It is understood that as the plasma density increases, i.e. as the inter-fermion distances decreases, the kinetic energy of free electrons may become comparable to their potential energy and they become trapped by ions (despite the high temperature values in compact-stars) and the plasma become strongly coupled possibly in a face-centered cubic or hexagonal lattice. In such condition plasma is said to be strongly coupled and therefore other considerations such as nonuniform electron distribution, Coulomb and exchange interaction effects must be taken into account in hydrodynamic plasma treatment. In fact it has been shown by Salpeter \cite{salpeter} that the compositional dependence due to the aforementioned effects in a white-dwarf may lead to minor corrections in mass-radius relation of Chandrasekhar. Furthermore, at densities greater than $10^9gr/cm^{3}$ which can be realized for the core of white-dwarfs, the effect of inverse $\beta$-decay (electron capture) has also to be invoked \cite{salpeter, hoyos}. Although, the electron relativistic-mass effect is usually ignored in the hydrodynamics study of quantum plasmas due to the dominating effect of high electron degeneracy pressure, however, this effect may become important in some extreme relativistic cases. Asenjo et al. \cite{asenjo} have worked out a complete hydrodynamics theory of relativistic quantum plasma including the electron spin effect.

In this work, using a generalized form of plasma pressure, we extend the standard quantum hydrodynamic (QHD) model to the extreme conditions where interaction effects are present. The presentation of the article is as follows. The extended QHD plasma model is introduced in Sec. \ref{equations}. Large amplitude localized solutions and the criteria of existence of such entities is derived in Sec. \ref{calculation}. The numerical analysis is presented in Sec. \ref{discussion} and a summary is given in Sec. \ref{conclusion}.

\section{Strongly Coupled QHD Model}\label{equations}

In this model, we consider a relativistically degenerate Fermi-Dirac electron-ion plasma with interacting electrons. Using the quantum hydrodynamics (QHD) basic set of equations in the center of mass frame, the continuity equation for the two-fluid plasma may be expressed as
\begin{equation}\label{cont}
\frac{{\partial \rho }}{{\partial t}} + \nabla \cdot(\rho {\bf{u}}) = 0,
\end{equation}
where, $\rho=m_i n_i + m_e n_e$ (using the neutrality $\rho\simeq m_i n_e$) and, $\bm{u}=(n_e m_e \bm{u_e} + n_i m_i \bm{u_i})/\rho$ are the center of mass density and velocity of the plasma. On the other hand, the momentum equation including the Bohm-force (electron tunneling effect) is given as \cite{haas1}
\begin{equation}\label{mom}
{\rho}\frac{{d{{\bf{u}}}}}{{dt}} =  - \nabla {P_{tot}} + \frac{{{\rho}{\hbar ^2}}}{{2{m_e}{m_i}}}\nabla \left( {\frac{{\Delta \sqrt {{\rho}} }}{{\sqrt {{\rho}} }}} \right),
\end{equation}
where, $P_{tot}=P_{int}+P_{non}$ is the total pressure due to interaction and non-interacting energies. The relativistic degeneracy pressure of non-interacting Fermi-Dirac gas is given by \cite{chandra1}
\begin{equation}\label{p}
P_{deg} = \frac{{\pi m_e^4{c^5}}}{{3{h^3}}}\left[ {R\left( {2{R^2} - 3} \right)\sqrt {1 + {R^2}}  + 3{{\sinh}^{ - 1}}R} \right],
\end{equation}
where, the relativity parameter, $R=P_{Fe}/(m_e c)=(\rho/\rho_c)^{1/3}$ ($\rho_c\simeq 1.97\times 10^6 gr/cm^3$) is a measure of the relativistic degeneracy of electrons and $P_{Fe}$ is the electron Fermi relativistic-momentum \cite{akbari7}.

On the other hand, the interaction pressure due to Coulomb attraction plus Thomas-Fermi nonuniform-distribution, electron exchange and ion-correlation effects are expressed in the following forms \cite{salpeter}, respectively
\begin{equation}\label{ctf}
{P_{C + TF}} =  - \frac{{8{\pi ^3}{m_e^4}{c^5}}}{{{h^3}}}\left[ {\frac{{\alpha {Z^{2/3}}}}{{10{\pi ^2}}}{{\left( {\frac{4}{{9\pi }}} \right)}^{1/3}}{R^4} + \frac{{162}}{{175}}\frac{{{{(\alpha {Z^{2/3}})}^2}}}{{9{\pi ^2}}}{{\left( {\frac{4}{{9\pi }}} \right)}^{2/3}}\frac{{{R^5}}}{{\sqrt {1 + {R^2}} }}} \right].
\end{equation}
where, $\alpha$($=e^2/\hbar c\simeq 1/137$) and $Z$ are the fine-structure constant and the atomic number, respectively.
\begin{equation}\label{ex}
\begin{array}{l}
{P_{ex}} =  - \frac{{2\alpha m_e^4{c^5}}}{{{h^3}}}\left\{ {\frac{1}{{32}}\left( {{\beta ^4} + {\beta ^{ - 4}}} \right) + \frac{1}{4}\left( {{\beta ^2} + {\beta ^{ - 2}}} \right) - \frac{3}{4}\left( {{\beta ^2} - {\beta ^{ - 2}}} \right)\ln \beta } \right. - \frac{9}{{16}} + \frac{3}{2}{\left( {\ln \beta } \right)^2} \\
\left. { - \frac{R}{3}\left( {1 + \frac{R}{{\sqrt {1 + {R^2}} }}} \right)\left[ {\frac{1}{8}\left( {{\beta ^3} - {\beta ^{ - 5}}} \right) - \frac{1}{4}\left( {\beta  - {\beta ^{ - 3}}} \right) - \frac{3}{2}\left( {\beta  + {\beta ^{ - 3}}} \right)\ln \beta  + \frac{{3\ln \beta }}{\beta }} \right]} \right\}. \\
\end{array}
\end{equation}
where, $\beta = R + \sqrt{1 + R^2}$.
\begin{equation}\label{cor}
{P_{cor}} =  - \frac{{311}}{{11250}}\frac{{\pi {\alpha ^2}{m_e^4}{c^5}}}{{{h^3}}}{R^3}.
\end{equation}
Taking the $x$-direction as the propagation direction, the dimensionless set of QHD equations is thus obtained using the following standard scalings
\begin{equation}\label{normal}
x \to \frac{{{c_{s}}}}{{{\omega _{pi}}}}\bar x,\hspace{3mm}t \to \frac{{\bar t}}{{{\omega _{pi}}}},\hspace{3mm}\rho \to \bar \rho{\rho_0},\hspace{3mm}u \to \bar u{c_{s}},
\end{equation}
where the normalizing factors, $\rho_0$, ${\omega _{pi}} = \sqrt {{4\pi e^2}{\rho_{0}}/{m_i^2}}$ and ${c_{s}} = c\sqrt {{m_e}/{m_i}}$ denote the equilibrium plasma mass-density, characteristic ion plasma frequency and ion quantum sound-speed, respectively. Also, the bar symbol over quantities denote the non-dimensionality and are avoided for simplicity in the forthcoming algebra. Thus, in terms of the effective quantum potential, $\Psi_{eq}$, we have the following closed set of QHD equations
\begin{equation}\label{dimensionless}
\begin{array}{l}
\frac{{\partial \rho}}{{\partial t}} + \frac{{\partial \rho{u}}}{{\partial x}} = 0, \\
\frac{{\partial {u}}}{{\partial t}} + {u}\frac{{\partial {u}}}{{\partial x}} = \frac{{\partial {\Psi _{eq}}}}{{\partial x}} + {H^2}\frac{\partial }{{\partial x}}\left( {\frac{1}{{\sqrt {\rho} }}\frac{{{\partial ^2}\sqrt {\rho} }}{{\partial {x^2}}}} \right), \\
{\Psi _{eq}} = \Psi_{deg}+\Psi_{C+TF}+\Psi_{ex}+\Psi_{cor}. \\
\end{array}
\end{equation}
We can obtain the quantum effective potential using the following relations
\begin{equation}\label{dimensional}
{\Psi _{eq}} =  - \int {\frac{{{d_R}{P_{tot}}(R)}}{R}} dR,\hspace{3mm}\frac{{\partial {\Psi _{eq}}}}{{\partial x}} = \frac{1}{\rho }\frac{{d{P_{tot}}(R)}}{{dR}}\frac{{\partial R}}{{\partial x}}.
\end{equation}
Thus, the effective potential in terms of the relativity parameter can be written in the following compact and normalized form
\begin{equation}\label{psi}
\begin{array}{l}
{\Psi _{eq}} = \sqrt {1 + R_0^2{\rho ^{2/3}}}  - 2\alpha \left( {\frac{{{6^{1/3}}}}{{175}}} \right){\left( {\frac{Z}{\pi }} \right)^{2/3}}\left[ {35{{(2\pi )}^{1/3}}{R_0}{\rho ^{1/3}} + \alpha {{(63{Z^2})}^{1/3}}\left( {\frac{{3 + 4R_0^2{\rho ^{2/3}}}}{{\sqrt {1 + R_0^2{{\bar \rho }^{2/3}}} }}} \right)} \right] +\\
\frac{\alpha }{{2\pi }}\left( {{R_0}{\rho ^{1/3}} - \frac{{3{{\sinh }^{ - 1}}({R_0}{\rho ^{1/3}})}}{{\sqrt {1 + R_0^2{\rho ^{2/3}}} }}} \right) - \frac{{311{\alpha ^2}\ln ({R_0}{\rho ^{1/3}})}}{{10000}}, \\
\end{array}
\end{equation}
in which, the normalizing relativistic degeneracy parameter $R_0=(\rho_0/\rho_c)^{1/3}$ and the Bohm tunneling-force parameter, $H = \sqrt {{m_i}/{2m_e}}(\hbar {\omega _{pi}})/({m_e}{c^2})$ are related through the simple relation $H = e\hbar \sqrt {{\rho_c}R_0^3/m_i\pi } /(2m_e^{3/2}{c^2})$. In Fig. 1 (left plot) a comparison has been made between the interaction effective-potentials and (right plot) the non-interaction one. It is observed that for a wide range of the relativity parameter, $R$, the correlation potential which is proportional to $\alpha^2$ is negligible compared to others, hence, will be ignored in forthcoming algebra.

\section{Relativistically Degenerate Density Excitations}\label{calculation}

In this section we will evaluate arbitrary-amplitude nonlinear density excitations in an interacting Fermi-Dirac plasma using the standard pseudopotential approach. To do so, we use the coordinate transformation $\xi=x-M t$ ($M$ being the Mach-number) which brings us to the co-moving frame reference, and rewrite the Eqs. (\ref{dimensionless}) in the new coordinate. We also use the appropriate boundary conditions, $\mathop {\lim }\limits_{\xi  \to  \pm \infty } \rho = 1$ and $\mathop {\lim }\limits_{\xi  \to  \pm \infty } u = 0$, to solve the continuity relation to give $u = M\left( {{1}/{\rho} - 1} \right)$. Furthermore, in order to calculate the Sagdeev pseudopotential we define a new variable, $\rho=A^2$ and use the value of $u$ together with the boundary conditions in the normalized momentum equation, Eqs. (\ref{dimensionless}) to find
\begin{equation}\label{diff2}
\begin{array}{l}
\frac{{{H^2}}}{A}\frac{{{d^2}A}}{{d\xi^2 }} = \frac{{{M^2}}}{2}{(1 - {A^{ - 2}})^2} - {M^2}(1 - {A^{ - 2}}) + \sqrt {1 + {R_0^2}{A^{4/3}}}  - \sqrt {1 + {R_0^2}}  \\- \alpha {\left( {\frac{{26{Z^2}}}{{175{\pi ^2}}}} \right)^{1/3}}\left[ {35{{(2\pi )}^{1/3}}R_0{A^{2/3}} + \frac{{{{63}^{1/3}}{Z^{2/3}}\alpha (3 + 4{R_0^2}{A^{4/3}})}}{{\sqrt {1 + {R_0^2}{A^{4/3}}} }}} \right] - \frac{\alpha }{{2\pi }}\left[ {R_0 - \frac{{3{{\sinh }^{ - 1}}R_0}}{{\sqrt {1 + {R_0^2}} }}} \right] \\ + \frac{{{{26}^{1/3}}{Z^{2/3}}\alpha }}{{175{\pi ^{2/3}}}}\left[ {35{{(2\pi )}^{1/3}}R_0 + \frac{{{{63}^{1/3}}(3 + 4{R_0^2}){Z^{2/3}}\alpha }}{{\sqrt {1 + {R_0^2}} }}} \right] + \frac{\alpha }{{2\pi }}\left[ {R_0{A^{2/3}} - \frac{{3{{\sinh }^{ - 1}}(R_0{A^{2/3}})}}{{\sqrt {1 + {R_0^2}{A^{4/3}}} }}} \right]. \\
\end{array}
\end{equation}
The above differential equation is of the form ${d^2}A/d{\xi ^2} = f(A)$ which can be multiplied by $dA/d\xi$ and integrated with respect to $\xi$ using the aforementioned boundary conditions in from $\xi=-\infty$ to $\xi=\infty$ to lead to the well-known energy integral of the form given below in terms of center of the mass density variable
\begin{equation}\label{energy}
{({d_\xi } \rho)^2}/2 + U(\rho) = 0,
\end{equation}
from which, after some algebra, the pseudopotential describing the nonlinear density excitations can be calculated as
\begin{equation}\label{pseudo}
\begin{array}{l}
U(\rho ) = \frac{1}{{700{H^2}\pi R_0^3\sqrt {1 + R_0^2} }}\left\{ {{R_0}\left[ {700{M^2}\pi R_0^2\sqrt {1 + R_0^2} {{(1 - \rho)}^2} + \rho } \right.} \right. \\
\times \left[ {{R_0}\alpha \left[ {1575\sqrt {1 + R_0^2} ({1 - \rho ^{2/3}})} \right.} \right. + 35R_0^2\sqrt {1 + R_0^2} \left( {{{43}^{1/3}}{{(2\pi )}^{2/3}}{Z^{2/3}} - 5} \right) \\ \times {({1 - \rho ^{1/3}})^2}(1 + 2{\rho ^{1/3}} + 3{\rho ^{2/3}}) + {963^{2/3}}{(2\pi )^{1/3}}R_0^3{Z^{4/3}}\alpha (1 - 4\rho ) + {2883^{2/3}} \\
\times {(2\pi )^{1/3}}{R_0}{Z^{4/3}}\alpha \rho \left. {\left( {\sqrt {1 + R_0^2} \sqrt {1 + R_0^2{\rho ^{2/3}}}  - 1} \right)} \right] + 175\pi \left[ {3 - 3\sqrt {1 + R_0^2} } \right. \\ \times \sqrt {1 + R_0^2{\rho ^{2/3}}} {\rho ^{1/3}} + R_0^4(8\rho  - 2) + {R_0^2} {\left. {\left. {\left. {\left( {1 + 8\rho  - 6\rho \sqrt {1 + {R_0^2}} \sqrt {1 + R_0^2{\rho ^{2/3}}} } \right)} \right]} \right]} \right]} \\ - 3\alpha \sqrt {1 + R_0^2} {({\sinh ^{ - 1}}{R_0})^2} - 525\rho \left[ {\left[ {\pi \sqrt {1 + R_0^2}  + 2{R_0}\alpha (3 + R_0^2 + 2R_0^2\rho )} \right]} \right. \\ \times {\sinh ^{ - 1}}{R_0} - \sqrt {1 + R_0^2} {\sinh ^{ - 1}}({R_0}{\rho ^{1/3}})\left[ {\pi  + 6{R_0}\alpha \sqrt {1 + R_0^2{\rho ^{2/3}}} {\rho ^{1/3}}} \right. \\ \left. {\left. {\left. { - 3\alpha {{\sinh }^{ - 1}}({R_0}{\rho ^{1/3}})} \right]} \right]} \right\}. \\
\end{array}
\end{equation}
The problem of localized nonlinear wave propagation for Sagdeev pseudopotential is quite analogous to a particle moving in a potential. In order to set a criteria for the existence of a localized density profiles we examine the following essential conditions to be satisfied all together in order for the pseudoparticle to be confined in a pseudowell, $U(\rho)$
\begin{equation}\label{conditions}
{\left. {U(\rho)} \right|_{\rho = 1}} = {\left. {\frac{{dU(\rho)}}{{d\rho}}} \right|_{\rho = 1}} = 0,\hspace{3mm}{\left. {\frac{{{d^2}U(\rho)}}{{d{\rho^2}}}} \right|_{\rho = 1}} < 0.
\end{equation}
It is further required that for at least one either maximum or minimum nonzero $\rho$-value, we have $U(\rho_{m})=0$, so that for every value of $\rho$ in the range ${\rho _m} > \rho  > 1$ (compressive soliton) or ${\rho _m} < \rho  < 1$ (rarefactive soliton), $U(\rho)$ is negative (it is understood that there is no other roots in the range $[1,\rho_m]$). In such a condition we can obtain a potential minimum which describes the possibility of a solitary wave propagation. The stationary soliton solutions corresponding to the pseudo-potential, $U(\rho)$, which satisfies the above-mentioned boundary-conditions, read as
\begin{equation}\label{soliton}
\xi  - {\xi _0} =  \pm \int_1^{\rho_m} {\frac{{d\rho}}{{\sqrt { - 2U(\rho)} }}}.
\end{equation}
It is observed that, the first condition for the existence of a solitary propagation is that it takes infinitely long pseudo-time ($\xi$) for the thermodynamic system to get away from the unstable point ($\rho=1$). The above statement is equal ti saying that $d_\rho U(\rho)\mid_{\rho=1}=0$ or equivalently $d_\xi \rho\mid_{\xi=-\infty}=0$ in parametric space which is also inferred by the shape of a solitary excitation. Secondly, moving forward in pseudo-time ($\xi$), the localized density perturbation reaches a maximum or a minimum at $\rho=\rho_m$ (i.e. the root if it exists) at which the pseudo-speed ($d_\xi \rho$) of the analogous particle bound in pseudo-potential, $U(\rho)$, in region $1>\rho>\rho_m$ (or $1<\rho<\rho_m$) reaches zero again and it returns back. Note also that parametrical speaking about equation Eq. (\ref{soliton}), $U(\rho)$ should be negative for solitary (non-periodic) wave solution, which is clearly satisfied if $d_{\rho\rho}U(\rho)\mid_{\rho=1}<0$ and $U(\rho_m\neq 1)=0$. Note also that, both the requirements $U(\rho)\mid_{\rho=1}=0$ and $d_{\rho}U(\rho)\mid_{\rho=1}=0$ follow from the equilibrium state assumption at infinite pseudo-time ($\xi=\pm\infty$) before and after perturbation takes place, i.e. $d_{\xi\xi} \rho\mid_{\xi=\pm\infty}=d_{\xi} \rho\mid_{\xi=\pm\infty}=0$.

It is easily conformed that, the pseudopotential given by Eq. (\ref{pseudo}) and its first derivative vanish at $\rho=1$, as required by the two first conditions in Eq. (\ref{conditions}). Also, the direct evaluation of the second derivative of the Sagdeev potential given by Eq. (\ref{pseudo}), at unstable density value, $\rho=1$, gives
\begin{equation}\label{dd}
\begin{array}{l}
{\left. {\frac{{{d^2}U(\rho)}}{{d{\rho^2}}}} \right|_{\rho = 1}} = \frac{1}{{525\pi {H^2}{{(1 + {R_0^2})}^{3/2}}}}\left\{ {R_0\alpha \left[ {70\sqrt {1 + {R_0^2}} \left( {5 + {{23}^{1/3}}{{(2\pi )}^{2/3}}{Z^{2/3}}} \right)} \right.} \right. \\ - 350\pi (1 + {R_0^2})\left( {{R_0^2} - 3{M^2}\sqrt {1 + {R_0^2}} } \right) + 35{R_0^2}\sqrt {1 + {R_0^2}} \left( {{{43}^{1/3}}{{(2\pi )}^{2/3}}{Z^{2/3}} - 5} \right) \\ \left. {\left. { + {{1203}^{2/3}}{{(2\pi )}^{1/3}}R_0{Z^{4/3}}\alpha  + {{963}^{2/3}}{{(2\pi )}^{1/3}}{R_0^3}{Z^{4/3}}\alpha } \right] - 525{R_0^2}\alpha \sinh R_0} \right\}, \\
\end{array}
\end{equation}
which defines an upper limit Mach-number value for existence of localized soliton-like density excitations. The value of this critical Mach-number is given below
\begin{equation}\label{consol2}
\begin{array}{l}
{M_{cr}} = \frac{1}{{5\sqrt {42\pi } \sqrt {{{(1 + R_0^2)}^{3/2}}} }}\left\{ {\left[ {{R_0}\left[ {350\pi {R_0}(1 + R_0^2) - \alpha \left[ {70\sqrt {1 + R_0^2} \left( {5 + {{23}^{1/3}}{{(2\pi )}^{2/3}}{Z^{2/3}}} \right)} \right.} \right.} \right.} \right. \\
\left. { + 35R_0^2\sqrt {1 + R_0^2} \left( {{{43}^{1/3}}{{(2\pi )}^{2/3}}{Z^{2/3}} - 5} \right) + {{1203}^{2/3}}{{(2\pi )}^{1/3}}R_0{Z^{4/3}}\alpha  + {{963}^{2/3}}{{(2\pi )}^{1/3}}R_0^3{Z^{4/3}}\alpha } \right] \\
\left. {{{\left. {\left. { + 525R_0\alpha \sinh {R_0}} \right]} \right]}^{1/2}}} \right\}. \\
\end{array}
\end{equation}
Furthermore, the examination of the limits for the pseudopotential leads to the following
\begin{equation}\label{nm}
\mathop {\lim }\limits_{\rho \to 0} U(\rho) = \frac{{{M^2}}}{{{H^2}}} > 0,\hspace{3mm}\mathop {\lim }\limits_{\rho \to \infty } U(\rho) = -\infty,
\end{equation}
revealing the fact that only rarefactive density profile is possible in this plasma. These limits along the third condition in Eq. (\ref{conditions}) grantee the existence of at least one root $\rho_m<1$ to exist.

\section{Numerical Interpretations}\label{discussion}

Figure 2 depicts the volume in $R_0$-$M$-$Z$ space in which the localized excitations are possible. It is remarked that, unlike for metallic densities, the dependence of the stability region on the nonuniform Thomas-Fermi electron distribution and Coulomb interaction corrections is insignificant in super dense plasmas even for very heavy-element composed ($Z=100$) Fermi-Dirac plasmas. This feature could be claimed as a proof for applicability of Fermi-Dirac electronic-gas plasma model previously employed for white-dwarf stars in literature \cite{akbari6, akbari8}. The rarefactive pseudopotential profiles shown (in Fig. 3) for this plasma for various plasma fractional parameters, reveal that in fact the change to the pseudopotential shape due to the change in the atomic-number is negligible for relativistic densities (Fig. 3(a)) compared to those of non-relativistic ones (Fig. 3(b)). However, the variation of the pseudopotential shape due to the change in the relativistic degeneracy parameter ($R_0$), despite the nature of its variation, is important for both relativistic (Fig. 3(c)) as well as non-relativistic (Fig. 3(d)) plasma density regimes.

The localized excitations corresponding to the values employed in Fig. 3 with similar variations are shown in Fig. 4. Figures 4(a) and 4(b) indicate that the localized density amplitude slightly decreases as the atomic number increases in the nonrelativistic degeneracy regime ($R_0<1$), while both its width and amplitude are almost invariant under this change in the relativistic degeneracy case ($R_0>1$). On the other hand, Figs. 4(c) and 4(d) reveal further distinct differences between the two degeneracy regimes when the atomic composition is fixed and the plasma fractional mass-density is varied. It is remarked that in the nonrelativistic degeneracy case the increase in mass density causes the density rarefaction (density depression) to widen and get deeper, while in the relativistic degeneracy case the increase in the plasma mass-density cause only the widening of the excitation gap without a change in its depth. These distinct features of Fermi-Dirac plasmas has already been reported in many recent articles \cite{akbari6, akbari9}.

\section{Concluding Remarks}\label{conclusion}

We have shown that distinct features of wave dynamics are present between the nonrelativistic and relativistic degeneracy limits in Fermi-Dirac plasmas even in the presence of Thomas-Fermi distribution, Coulomb and exchange interactions, and correlation effects. Furthermore, it was revealed that the crystallization and other corrections which are proportional to the fine-structure constant become dominant for nonrelativistic plasma densities, but become ineffective for relativistic densities where the electron quantum tunneling and degeneracy pressure effects prevail over the minor corrections. This feature is specific to the astrophysical compact plasma densities and is opposite to the case of ordinary solid state densities observed in the laboratory. Hence, the previous simplified QHD Fermi-gas model application to relativistically degenerate white dwarf stars is justified.

\newpage

\textbf{FIGURE CAPTIONS}

\bigskip

Figure-1

\bigskip

The appropriately scaled effective interacting quantum potentials due to Thomas-Fermi plus Coulomb, exchange and correlation pressures are compared in the left plot. The right plot showing the scaled effective non-interacting quantum pressure due to Chandrasekhar relativistic degeneracy pressure.

\bigskip

Figure-2

\bigskip

Figure 2 shows a volume in $M$-$Z$-$R_0$ space in which a localized density excitation can exist.

\bigskip

Figure-3

\bigskip

(Color online) The variations of pseudopotential depth and width for rarefactive localized nonlinear density waves with respect to change in each of three independent plasma fractional parameter, normalized soliton-speed, $M$, the atomic number, $Z$, and the relativistic degeneracy parameter, $R_0$, while the other two parameters are fixed. The dash-size of the curve increases according to increase in the varied parameter.

Figure-4

\bigskip

(Color online) The variations of void-type localized density-structure with respect to change in each of three independent plasma fractional parameter, normalized soliton-speed, $M$, the atomic number, $Z$, and the relativistic degeneracy parameter, $R_0$, while the other two parameters are fixed. The thickness of the solitary profile increases according to increase in the varied parameter.

\bigskip

\end{document}